# EFFECTIVE QUERY RETRIEVAL SYSTEM IN MOBILE BUSINESS ENVIRONMENT

## [1]R.Sivaraman, [2] RM.Chandrasekaran

[1]Dy.Director, Center for Convergence of Technologies (CCT), Anna University Tiruchirappalli, Tiruchirappalli, Tamil Nadu, India

email: rsiva.raman@yahoo.com

[2]Registrar, Anna University Tiruchirappalli, Tiruchirappalli, Tamil Nadu, India

Email: aurmc@sify.com

*Abstract*

Web Based Query Management System (WBQMS) is a methodology to design and to implement Mobile Business, in which a server is the gateway to connect databases with clients which sends requests and receives responses in a distributive manner. The gateway, which communicates with mobile phone via GSM Modem, receives the coded queries from users and sends packed results back. The software which communicates with the gateway system via SHORT MESSAGE, packs users' requests, IDs and codes, and sends the package to the gateway; then interprets the packed data for the users to read on a page of GUI. Whenever and wherever they are, the customer can query the information by sending messages through the client device which may be mobile phone or PC. The mobile clients can get the appropriate services through the mobile business architecture in distributed environment. The messages are secured through the client side encoding mechanism to avoid the intruders. The gateway system is programmed by Java, while the software at clients by J2ME and the database is created by Oracle for reliable and interoperable services.

## I. INTRODUCTION

Due to the growth of Mobile network and data management schemes, mobile business has drawn attention by more customers in distributive environment. As a result, it must determine how to deliver compelling applications to ensure that data services to fulfill their potential requirements.

In order to provide mobile business applications to the consumers, it is necessary to store and retrieve persistent information on the mobile device as well as access remote information stored on a remote DBMS host from the mobile device. The storage of information on the mobile device is necessary because a business application can stop suddenly due to different reasons, such as, an incoming call phone or because the device runs out of battery.

As of now, most of the web based query management schemes are in preliminary stage only, and the goal of this query system is to achieve that can be achieved by PC but now by mobile phones.

With an short message based query message, data could be retrieved by customers in mobile business environment and unnecessary time would not be wasted. Due to the reliable nature of short message, the consumers will receive the message even if their phone is turned off at that time.

The short message in web based Query management system described in this paper is the interaction short message, which gives only the requested information.

## II. RELATED WORK

As of now, most of the existing web based business query management scheme is in preliminary stage only, for example, a small number of query via short message are encoded in PDU (Packet Data Unit) format, which will increase the cost when the payload increases and it also leads to sending more than one message if the requested payload is large. So the traditional PDU method is not suitable for mobile business queries in web based distributed environment.

## III. PROPOSED BUSINESS QUERY STRUCTURE

Web based mobile business query structure proposed here is differed from the traditional PDU format, uses request and response short message, which makes use of short message to achieve two-way delivery of information. Message of this business gateway system, which in special data structure, is encoded in Text mode based on AT commands as compared to PDU format in traditional method, and that will greatly shorten the length of short message since the message is coded into a standard format.

The structure of the mobile business Querying System for Academic Education which is composed of mobile phone, data source server (Business Data Management) and the gateway, is given below.

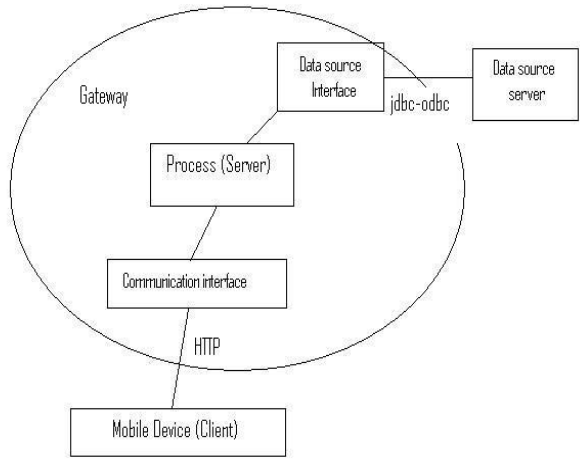

Fig.1 Structure of mobile business query system

The application software which communicates with the gateway system via short message connection packs the query code, User ID and Password and sends encoded package to the gateway. The gateway will receive the coded queries from customers, retrieve the data according to the query code and sends packed results back. The software at client side, then interprets the packed data for the users to read on a page of GUI. The system can be extended to support GPRS, CDMA, Bluetooth and other communication protocols.

Gateway must be setup in such a manner as shown in Fig. 1 to establish short message connection between mobile phones and data management system host. The gateway is composed of Communication Interface Module, Processing Module and Data Source Interface Module. This work takes educational query as an example to analyze the implementation of web based mobile business query system.

## IV. THE BUSINESS GATEWAY SYSTEM

Mobile business Query based on short message is achieved by the connection between the server, SHORT MESSAGE and database management system.

When business application installed in mobile phones is initiated, and then user has to select any one business query from the available choices and sends the packed query information in the form of short message to the gateway system. Communication interface module which is connected serially to the Processing module will receive the short message and gives it to the processing module. Processing module decodes the received business query message and fetches the user's query information (query code, User ID, and Password) and first checks whether the user is valid to query the business information or not. Processing module then converts the query information to SQL and queries the database according to the business query code. Finally the results which are encoded into short message are sent back to the user's destination number through short message. The software application in mobile device displays the results in the form of text. That's called a complete query.

### A. Communication Interface Module

Communication Interface Module establishes a communication channel between mobile phones and communication module via short message connection. A database is created in communication interface module to store the information of short messages such as user's phone number, user ID, Date/Time, query code etc. The Database also has four tables to increase the reliability of the mobile business query system, they are, InBox table, OutBox table, BackUp table and FailureSend table.

### 1) InBox Table:

InBox table is used store the received business query messages from the serial communication port before processing that short messages. Processing module, which is always monitoring InBox table and fetches one message at each time when there are messages in this table.

### 2) OutBox Table:

OutBox table stores the messages that are waiting for being sent to the corresponding mobile phone numbers. Once the message has been sent, it is cleared from the OutBox table and starts sending the next message from the table.

### 3) BackUp Table:

When processing module has finished the query process of short messages, then processing module clears the short message from InBox table and preserves it to the BackUp table for future use. All short messages in BackUp table will be removed every week.

### 4) FailureSend Table:

A business application can stop suddenly at any point of time. FailureSend table stores messages that didn't be sent due to some failures like network failure, out of range and battery failure etc., Processing Module checks whether there is a message in this table after a particular period of time if yes then fetches the message(s) and sends it into OutBox table after re-set up.

*B. Back end (Web) Module*

The Processing Module, which is responsible to provide the interface between the communication interface module and data management server together and ensures the complete implementation of the mobile business query function. Once the system is started, web module of the business system monitors the InBox Table and it deals with one message each time according to First Come First Serve (FCFS) basis. This module fetches sender's number and sends it to ReceiverMobileNo field of OutBox table, then fetches the contents from the message which includes three data's for further processing: query content, student's ID number or teacher's number, and password.

First Web Module authenticate whether the user is legal according to user's ID and password using SHORT MESSAGE database which is extracted from the message, then determines which database should be queried after integrates query information, –query content and additional element, and generates SQL to query database, finally puts the message into OutBox table of Communication Interface Module waiting for being sent. Progressing Module takes multi-thread technique to judge the query.

Since customer's request message arrives at a random time, there are a number of messages arrive within a certain period of time. Moreover, the procedure of processing every customer's query message from querying database to sending messages by GSM Modem should be safe and reliable. Consequently multi-thread technique is exerted in the whole program. Main thread creates a new sub-thread for each query process when there is a new message. The sub-thread processes the new message and won't be released until the result is sent back to user.

*C. Data Management Module*

The data management server which has been created using oracle for education department includes: students' score table, students' credit table, students' exam schedule table, teachers' information table, teachers' schedule table. Inquiry of business is carried out by connecting to the database server and table which has been confirmed according to query content extracted from the message.

## V.  THE BUSINESS SOFTWARE IMPLEMENTATION

As short messaging services is based on store and forward mechanism, the transportation of messages is safe and reliable. The business query system designed in this paper combines both Advanced User Interface and Low-level User Interface to a portable, extensible component of GUI on account of different mobile terminals and functions.

The application software of the mobile business is composed of four layers; they are user interface, record management system (RMS), the application software and Embedded OS.

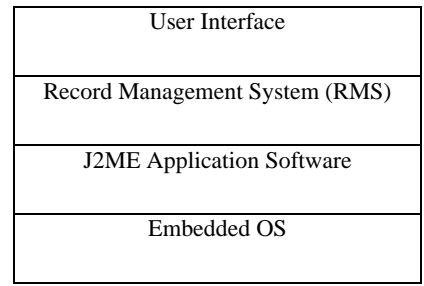

| User Interface |
| Record Management System (RMS) |
| J2ME Application Software |
| Embedded OS |

Fig.2 Layers of Business Software Implementation

*A. Sending Module*

Two forms are created using J2ME technology and is visualized by customers using Graphical User Interface (GUI) which is provided by sending module. First a Form is designed to provide query content for users to select from the list available choices. When users finished their choices, click "OK" button to go into another Form in which there are two TextFields for users to input their user IDs and passwords. This module then transforms query content, user ID and password into digital code after users finished selecting the choices. This can greatly shorten the length of short message. The sending module starts a new thread when user clicks the "send" button to avoid no response due to network blocking, no coverage etc.,

First, short message connection is established through Connector.open(url) method of message connection, in which the connection between the client mode and appropriate port number are set. Host number of destination address is the SIM card number of the GSM modem and destination address is indicated in sending module program.

*B. Receiving Module*

Receiving the result message of business query takes use of Push mechanism. After the introduction of MIDP2.0, Midlet applications can startup asynchronously through network connection.

A Midlet application of receiving messages is registered into a Push Registry through a push event by means of static registration. When there is a new message arrives the application manager creates a new instance of receiving message by new() method and invokes startApp() method to activate this Midlet. ListConnection() method is used to establish Message connection and open() method is invoked to open and process message. The connection will close after the process is finished.

As soon as the application starts, the receiving module monitors the monitors the designated port, then starts a new thread to receive message when there is a new message, and displays it by translating the message into text form user can read.

### C. Record Management System (RMS)

Persistent storage of result message is helpful for future use. The "javax.microedition.rms" package is based on Record Stores. A Record Store is the equivalent of a simple file. A Record Store stores a set of records in binary format. This package allows developers to store and retrieve information to/from files on mobile devices. Two Record Stores are designed in this paper, one is CourseStore which is used to store the code and name of courses, the other is ClassroomStore which is used to store the code and name of teaching buildings. When curriculums and teaching buildings need to be displayed, the application can access them at any time.

### VI. SHORT MESSAGE SPECIFICATION

The Short Message used for web based business query system contains three parts includes code of the query content, User ID and Password. The contents of short message are listed below in Table 1.

TABLE I
CONTENT OF SHORT MESSAGE

| Query Content | Code | User ID | Password |
|---|---|---|---|
| Student's score | 001 | Student's Register number (or) Teacher's Employee number | ------- |
| Student's credit | 002 | | |
| Teacher's Quantity | 003 | | |
| Teacher's class Quality | 004 | | |
| Exam schedule | 005 | | |

The 1st, 2nd and 3 rd characters are the reference which can help decide to query which database. The 4th to 15th characters are on behalf of students' number or teachers' number. The 18th to 27th characters are user's password.

When query code is '002', for example, it means the Attendance credit of a student is considered to be queried.

The objective of the paper is to reduce the length of the short message in this query system. In order to achieve this each part is assigned to a particular length of bits (digital codes) which will undoubtedly reduce the length and cost of the short message based query system. Since, the message is encoded into a standard format.

The length of the each part of Short Message in this system is listed in Table 2.

TABLE II
LENGTH OF SHORT MESSAGE

| Code of Query Content | User ID | Password |
|---|---|---|
| 3 bits | 12 bits | 10 bits |

### VII. CONCLUSION & FUTURE SCOPE

This project application starts with welcome page with time interval set to 2000 milliseconds before it will load the options form to provide a list of choices. The login form will then be displayed. This login form requires users to enter their login name and password. Only valid users can pass through this form before they can proceed to other forms. If the login name and password are not valid, error message will be displayed in the user's mobile devices.

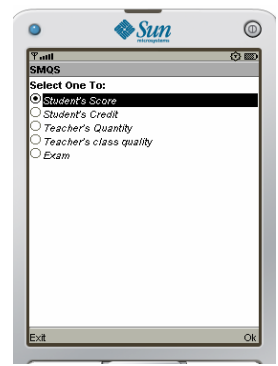
Fig.3 Form I

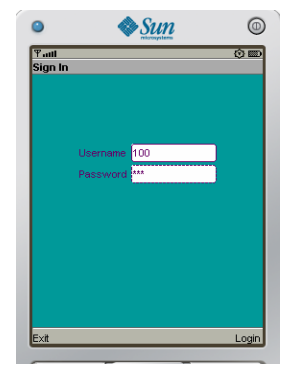
Fig. 4 Form II

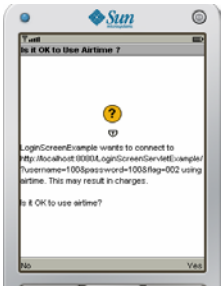
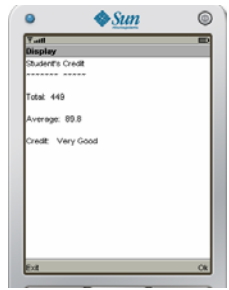

Fig. 5 Interface                    Fig.6 Result display

The method to implement mobile business query system based on request-response short message was described and developed in this work. This system is software designed based on typical practical applications to be used by business providers and customers. Even if there are several businesses only one query message according to a particular content query can be sent and the result is displayed correct. This designed software was found to be reliable and practical. It is verified that the system is usable and easy-operating.

In Future work, more business query functions can be added from the prototype design to achieve Book query, Price query, shopping query and other business queries. When there are a large number of short messages waiting for processing, GSM Modem group can be used to solve this problem to increase the reliability of the system.